\def\vep{{\cal E}}
\def\ba{\beta}
\def\ep{\epsilon}
\def\om{\omega}
\def\Om{\Omega}
\def\a{\alpha}

\def\pa{\partial}
\def\vp{\varphi}
\def\th{\theta}
\def\f#1#2{{#1\over#2}}

\documentstyle[12pt]{article}
\begin{document}
\begin{titlepage}
\begin{flushright}
gr-qc/9505025
\end{flushright}
\vfill
\begin{center}
{\Large\bf Orbits, forces and accretion dynamics near spinning black
holes}\\
\vfill
{\large\bf C. Barrab\`es,{$^{1,2}$}
 B. Boisseau,$^1$ and W. Israel$^3$}
\vfill
 {\small $^1$ Mod\`eles de Physique Mathematique,\\ EP 93 CNRS,
Universit\'e de Tours, F-37200, Tours (France)\\
$^2$ D\'epartement d'Astrophysique Relativiste et
Cosmologie, CNRS,\\ Observatoire de Meudon, F-92195, Meudon (France)\\
$^3$ Canadian Institute for Advanced Research Cosmology
Program,\\ Theoretical Physics Institute, University of Alberta,
Edmonton, Canada T6G 2J1}
\end{center}
\vfill
\normalsize
\abstract{We analyze the relativistic dynamical properties of Keplerian and
non-Keplerian circular orbits in a general axisymmetric and stationary
gravitational field, and discuss the implications for the stability
of co- and counter-rotating accretion disks and tori surrounding
a spinning black hole. Close to the horizon there are orbital
peculiarities which can seem counterintuitive, but are elucidated
by formulating the dynamics in terms of the orbital velocity
actually measured by a local, zero-angular-momentum observer.}
\vfill
\centerline{Keywords: accretion, \uppercase{AGN} and quasar models}
\vfill
\end{titlepage}

\baselineskip 15pt
\section{Introduction}

Accretion onto supermassive black holes has emerged as the
paradigmatic model for the energy source of AGNs and quasars
(Lynden-Bell 1978, Rees 1984), a picture recently given compelling support by
HST spectrography of the nucleus of M87 (Harms et al. 1994).

Detailed development and understanding of such models requires
insight into the subtle fluid and plasma dynamics close to the edge
of a spinning black hole. For near-maximal spin, accretion disks can
extend almost down to the horizon (modulo the possibility of still
poorly understood bar-like instabilities, see Blaes 1987). In this
strong-gravity regime, general-relativistic corrections can produce
non-Newtonian behaviour which, at first sight, is strange and
counterintuitive (see Allen 1990). For instance, angular momentum is
transported inwards rather than outwards by a viscous disk (Anderson
and Lemos 1978); there is a reversal of the Rayleigh criterion for
stability (which conventionally requires the specific angular
momentum to increase outwards) (Abramowicz and Prasanna 1990); and
spinning balls of fluid undergoing slow collapse actually
become more spherical (rather than more flattened) in
the latter stages of contraction toward black holes (Chandrasekhar
and Miller 1974).

An insightful series of papers by Abramowicz and co-workers
(Abramowicz and Lasota 1986, Abramowicz 1990, Abramowicz and Prasanna
1990) reveal how these seemingly disparate anomalies can be
understood in terms of a common, non-Newtonian mechanism: an
effective reversal of centrifugal force inside a region called the
rotosphere, which lies within radius $r=3 m$ for a
Schwarzschild black hole. (This line of thought is further pursued
and expounded in Abramowicz 1992, 1993 and Abramowicz and Szukiewicz
1993.)

Viewing a thing from more than one angle often adds depth and
flexibility to one's understanding. The group of anomalies that fall
under the heading ``Abramowicz effect'' can also be ``explained''
without overturning familiar Newtonian preconceptions (de~Felice
1991, 1994; Page 1993). Indeed, one very simple explanation, which we shall
present in this article, actually exploits the close formal {\it
resemblance\/} between the Newtonian and general-relativistic
descriptions (when suitably formulated). The basic remark is that the
general relativistic formulae become simplest and almost
indistinguishable from Newtonian when expressed in terms of
quantities that a local observer actually measures.

The gist of the matter can be explained in a few words. The outward
rocket thrust needed to hold a spacecraft in a circular orbit of
radius $r$ around a central mass is proportional to $(v_K^2-v^2)/r$
according to Newtonian theory, where $v$ and $v_K$ are the actual and
Keplerian orbital velocities. {\it This remains true in Einstein's
theory}, for a spherisymmetric field, if $v$, $v_K$ are interpreted
as velocities measured by a local stationary observer.

This at once makes it clear why, within $r=3m$ (the orbital
radius of a circling photon in the Schwarzschild field), the thrust
needed is always outward, no matter how fast the spacecraft orbits,
because here the Keplerian velocity is tachyonic: $|v_K|>c$. This is
the essence of the Abramowicz effect.

Moreover, this way of looking at the phenomenon carries over
trivially to the axisymmetric case when the central mass is in
(steady) rotation. (By contrast, on other interpretations
this extension is less straightforward (Abramowicz, Nurowski and Wex
1993).)
The thrust needed to stay on an equatorial circle is now proportional
to $(v_+-v)(|v_-|+v)$, where $v_+$ and $v_-$ are the prograde and
retrograde Keplerian velocities. All velocities are understood to be
measured by a local observer (``ZAMO'') orbiting with zero angular
momentum. Because of rotational frame-dragging, the two Keplerian
velocities are not equal and opposite; they become lightlike at
different radii, the retrograde orbit at the larger radius.

Such effects of frame-dragging on counter-rotating orbits have an
interest which may be more than merely one of principle. Recent
observations and numerical simulations suggest that counter-rotation
may be a less unusual feature in old galactic nuclei than was thought
just a few years ago (Binney and Tremaine 1987). The gas stream
produced by a retrograde encounter between galaxies can evolve over
billions of years into a counter-rotating disk of stars superimposed
upon and streaming collisionlessly through the old disk. NGC4550 and
NGC7217 are examples of such ``two-way galaxies'' (Rubin et al. 1992,
Kuijken 1993).

In Sec.~2 we introduce the basic concepts and techniques in the
simple context of a spherically symmetric gravitational field. This
introductory discussion clears the way for a concise presentation of
the corresponding results for equatorial circular orbits in a general
stationary axisymmetric field in Sec.~3. When the mass of the
accretion disk or torus is small compared to the central black hole,
the gravitational field is well approximated by the Kerr metric. In
this case, explicit results are available, and are summarized in
Sec.~4. Further developments and details are assembled in two
Appendices.

\section{Spherisymmetric fields}

We begin  by considering orbits in a general static, spheri\-
symmetric
geometry, described by the line-element
$$
ds^2=\eta^2(r)\,dr^2+r^2\,d\Om^2-V^2(r)\,dt^2.
$$

Without essential loss of generality we may assume the orbit confined
to the equatorial plane $\th=\pi/2$. We write $\om=\dot\vp\equiv
d\vp/dr$ for the angular velocity as measured by a comoring observer,
using proper time $\tau$.

Keplerian orbits are timelike geodesics, i.e., extremals of the
action $\int L\,d\tau$, where
$$
L=V^2 \dot t^2-r^2\om^2-\eta^2\dot r^2
$$
and $L=1$ on the extremal curve. For a circular Keplerian orbit of
radius $r$, the angular velocity $\om_K$ is thus found from $\pa
L/\pa r=0$ to be given by
\begin{equation}
\f{r\om_K^2}{1+r^2\om_K^2}=\f{V'(r)}V.
\end{equation}

With respect to a local stationary observer, a unit test mass on a
circular orbit has 3-momentum (i.e., spatial projection of 4-momentum
$p^\mu=dx^\mu/d\tau$)
$$
p=(g_{\vp\vp})^{\f12}\,p^\vp=r\om
$$
and 3-velocity
\begin{equation}
v=p/(1+p^2)^{\f12}=r\om/(1+r^2\om^2)^{\f12}
\end{equation}
in relativistic units $(c=1)$.

Hence (1) yields
\begin{equation}
v_K^2/r=V'(r)/V(r)
\end{equation}
for the Keplerian 3-velocity $v_K$. (In the Newtonian limit $V\approx
1+V_{{\rm Newt}}/c^2$, (1) and (3) reduce to the expected familiar
results.)

The orbital radius $r=r_{ph}$ of a freely circulating photon
$(|v_K|=1)$ is determined by
$$
1/r_{ph}=V'(r_{ph})/V(r_{ph}).
$$
In the case of the Schwarzschild geometry, $V^2=1-2m/r$ and
$r_{ph}=3m$. Inside this radius, Keplerian velocities are
tachyonic $(v_K^2>1)$, corresponding to spacelike geodesic
trajectories, $d\tau_K^2<0$.

We now turn to general (non-Keplerian) circular orbits. The
mechanical force or rocket thrust needed to hold a unit test mass
stationary in such an orbit is given by $F^\mu=\delta
p^\mu/\delta\tau$ (the absolute derivative of 4-momentum), a purely
spatial and purely radial vector. Its radial component is
$$
F^r=\f{\delta^2r}{\delta
\tau^2}=-\f12(g_{rr})^{-1}\bigg\{\f{d}{d\tau}\bigg(\f{\pa L}{\pa \dot
r}\bigg)-\f{\pa L}{\pa r}\bigg\}=\f12 \eta^{-2}\,\f{\pa L}{\pa r}.
$$
This yields for the (signed) magnitude of (outward) force
\begin{equation}
F=(g_{rr})^{\f12} F^r=\f{V'}{\eta V}\bigg\{1-\om^2
r\left(\f{V}{V'}-r\bigg)\right\}.
\end{equation}
Recalling (1), this can be re-expressed as
\begin{equation}
F=\f{V'}{\eta V}\left(1-\f{\om^2}{\om_K^2}\right).
\end{equation}

Setting $\om=0$, we obtain the force $F_{{\rm stat}}=V'/\eta V$
needed to hold up a stationary unit mass. At the photon radius
$r=r_{ph}$, we have $\om_K^2=(d\vp/d\tau)_K^2=\infty$, hence
$F=F_{{\rm stat}}$ is finite for every timelike circular orbit and
velocity-independent (Abramowicz and Lasota 1986).

The subtracted term in (5), proportional to $\om^2$, might be thought
of as ``centrifugal force.'' Were we to adopt this terminology, we
would at once make contact with Abramowicz 1990. In the
rotosphere, Keplerian circular orbits are spacelike, hence
$\om_K^2<0$ in (5) and ``centrifugal force is reversed.''

But there is an alternative way to write (5). Using (2) to eliminate
$\om$ in (4) in favour of $v$, and recalling (3), gives
\begin{equation}
F=\f1{\eta
r}\,\f{v_K^2-v^2}{1-v^2}.
\end{equation}
In this form, ``centrifugal force'' (now most naturally interpreted
as the subtracted term $v^2/\eta r(1-v^2)$ in (6)) never changes
direction, since $(1-v^2)$ is positive for every physical (timelike)
orbit. Nevertheless, (6) shows as before that to orbit in the
rotosphere (where $v_K^2>1$) always requires an outward thrust $F$,
no matter how fast one moves. Indeed, speeding up is self-defeating:
the thrust needed actually increases with speed in the rotosphere and
becomes infinite at the speed of light.
Within the rotosphere, it takes less outward thrust to
hold a body stationary than to maintain it in orbit.
One might say that the
special-relativistic increase of ``mass'' with velocity (given by the
denominator of (6)) literally outweighs the effect of centrifugal
force in the rotosphere (Page 1993, de~Felice 1991).

It is this property that underlies the reversal of the Rayleigh
criterion for stability in the rotosphere. Consider a toroidal
distribution of material (e.g. incompressible fluid) held in steady,
differential rotation around a gravitating mass by a prescribed
non-gravitational force field. Suppose the specific angular momentum
of the distribution {\it increases\/} outwards from the axis. An
orbiting ring of material, if displaced inwards axisymmetrically,
will preserve its angular momentum and will therefore orbit faster
than its new surroundings. In the rotosphere, the outward force that
would be called for to support it in this new orbit is therefore
necessarily larger than that which actually supports the surrounding
material, and this is as much as the local force field provides. The
displaced ring is forced to sink further, and this triggers an
instability.

The alternative ``explanations'' of Abramowicz and Page, however
helpful and suggestive, are, of course, at bottom merely different
forms of words to clothe the formulae. One may adopt either (or
neither), depending on circumstances and personal taste. One
advantage of Page's form, as we shall now see, is that it is easy to
adapt (6) (but less easily (5)) to the case where the central mass is
rotating.

\section{Stationary axisymmetric fields}

We pass now to orbits in a stationary axisymmetric asymptotically
flat geometry. Let
$$
\xi_{(t)}^\a=\pa x^\a/\pa t,\qquad \xi_{(\vp)}^\a=\pa x^\a/\pa\vp
$$
be the timelike and axial Killing vectors (assumed to commute). An
observer with 4-velocity $u^\a=dx^\a/d\tau$ has angular momentum
$$
\ell=p_\vp=u\cdot \xi_{(\vp)}=g_{\vp \a}u^\a.
$$
For a free-falling (geodesic) observer, this is conserved.

Any observer orbiting in an azimuthal circle has a 4-velocity of the
form
\begin{equation}
u^\a=U^{-1}(\xi_{(t)}^\a+\Om\,\xi_{(\vp)}^\a),
\end{equation}
where $\Om=d\vp/dt$ is his angular velocity as measured by a
stationary observer at infinity. The orbiting observer is a zero
angular momentum (ZAM) observer (Bardeen 1970\thinspace a, 1973,
Thorne, Price \& MacDonald 1986) if
$u\cdot\xi_{(\vp)}=0$, i.e., if he orbits with the Bardeen angular
velocity
$$
\Om_B=-(\xi_{(\vp)\cdot}\,\xi_{(t)})\big/
(\xi_{(\vp)\cdot}\,\xi_{(\vp)})=-g_{\vp t}/g_{\vp\vp}.
$$
The normalizing factor in (7) is then
$$
U_{\rm ZAM} =V=(-g^{tt})^{-\f12}.
$$

The line-element is now conveniently expressible as
\begin{equation}
ds^2=g_{AB}
dx^A\,dx^B+\rho^2\,\sin^2\theta(d\vp-\Om_B\,dt)^2-V^2\,dt^2
\end{equation}
where $x^A\equiv (x^1,x^2)=(r,\theta)$ and all metric coefficients
depend on $r$ and $\theta$ only.

We assume that this geometry also has equatorial symmetry, and we
shall be interested in circular orbits in the equatorial plane
$\theta=\pi/2$. As in the previous section, their properties are
derivable from the Lagrangian
$$
L=V^2 \dot t^2-\rho^2\,\om^2-\eta^2\dot r^2
$$
in which the dot denotes differentiation with respect to proper time
$\tau$,
$$
\om=\dot\vp-\Om_B \dot t
$$
is the locally measured angular velocity relative to the local ZAM
observer and all coefficients are functions of $r$ only. The
3-velocity $v$ of a circularly orbiting particle as measured by the
local ZAM observer is
$$
v=\rho\om/(1+\rho^2\om^2)^{\f12}.
$$

Keplerian circular orbits satisfy $\pa L/\pa r=0$, $L=1$, $\dot r=0$.
This yields a quadratic equation with roots
\begin{equation}
v_+=\gamma/(\delta+\beta),\qquad
v_-=-\gamma/(\delta-\beta)
\end{equation}
giving the 3-velocities of prograde and retrograde Keplerian orbits.
We have defined
\begin{eqnarray}
\a=\pa_r \ln \rho,\qquad \beta&=-\f12(\rho/V)\pa_r\Om_B,\qquad
\gamma=\pa_r \ln V,\nonumber \\
\delta&=(\a\gamma+\beta^2)^{\f12}.
\end{eqnarray}
The retrograde Keplerian orbit becomes lightlike when
\begin{equation}
v_-=-1\ \Longrightarrow\ \gamma-\a=-2\beta,\quad
\delta=\f12(\a+\gamma).
\end{equation}
For the prograde orbit,
\begin{equation}
v_+=+1\ \Longrightarrow\ \gamma-\a=2\beta,\quad
\delta=\f12(\a+\gamma).
\end{equation}

The thrust required to hold unit mass in a non-Keplerian circular
orbit is given by the absolute derivative $F^\mu=\delta^2
x^\mu/\delta\tau^2$. The only nonvanishing component is calculable
from
$$
F^r=\delta^2 r/\delta\tau^2=\f12 \eta^{-2} \pa L/\pa r.
$$
This gives for the (signed) magnitude of outward thrust
$F=(g_{rr})^{\f12} F^r$:
\begin{equation}
F=-\f{\a}{\eta}\,\f{(v-v_+)(v-v_-)}{1-v^2}.
\end{equation}

Equations~(9) and (13) are invariant, as they should be, under
reparametrization of the arbitrary radial co-ordinate $r$. A more
obviously geometrical form emerges if we introduce the element of
proper radial distance
\begin{equation}
ds_r=(g_{rr})^{\f12}\,dr.
\end{equation}
Then
\begin{equation}
F=-\f1\rho\,\f{(v-v_+)(v-v_-)}{1-v^2}\,\f{d\rho}{ds_r}.
\end{equation}
The conical deficit factor $d\rho/ds_r$ (slope of circumferential
radius $\rho$ versus proper radial distance) is generally not far
from unity except very close to a black hole horizon, where it tends
to zero.

Setting $v=0$ in (13), and using (9) and (14), one obtains the
outward force needed to hold unit mass in a ZAM orbit:
$$
F_{\rm ZAM}=d\ln V/ds_r.
$$
This generalizes to non-equatorial ZAM orbits: its covariant form is
\begin{equation}
(F_\a)_{\rm ZAM}=u_{\a|\beta}\,u^\beta=\pa_\a\ln V.
\end{equation}
(A simple derivation is sketched in Appendix~A.)

ZAM photons emitted outward from deep in the field reach infinity
redshifted by a factor $V^{-1}$. At a black hole horizon, $V$ becomes
zero, the ZAM orbits become the horizon's lightlike generators and
the ``redshifted force'' $VF_{\rm ZAM}$ becomes the surface gravity
$\kappa$. Equation~(16) and these useful properties define the sense
in which $V\equiv (-g^{tt})^{-\f12}$ functions as the scalar
potential appropriate to ZAM observers (see, e.g., Israel 1983).

For definiteness, we take the angular momentum of the hole (or other
central mass) to be positive. Then ZAM observers are dragged in the
positive-$\vp$ direction ($\Om_B>0$), at least near the source. The
inequalities
\begin{equation}
\a>0,\qquad \ba>0,\qquad \gamma>0
\end{equation}
must hold, at least if the source is isolated. According to (10),
these  just express the conditions (respectively) that, as one moves
out from the source, circumferences of equatorial circles get larger,
effects of frame-dragging (as measured by $\Om_B$) get progressively
smaller, and
that ZAM observers always experience inward gravitational pull. The
inequalities (17) should remain valid for a larger class of
(non-isolated) sources, in particular those for which a surrounding
accretion disk or torus has a mass appreciably less than the central
mass.

Assuming the validity of (16), we infer at once from (9) that
$$
|v_-|>v_+,
$$
showing that, at a radius $r_{-ph}$ where retrograde Keplerian
orbits become lightlike, prograde Keplerian orbits must still exist.
Thus, quite generally there is an ``outer rotosphere,'' characterized
by $v_-<-1$, $v_+<1$, in which retrograde Keplerian orbits (only)
have become superluminal, and in consequence the Rayleigh criterion
is reversed for counter-rotating disks and tori.

Retrograde orbits can still extend into the outer rotosphere if they
are supported, for instance by the pressure gradient in a thick disk.
Counter-rotation in the outer parts of such a disk must, however,
give way to co-rotation in the inner parts, since
no outward radial force is available to hold up the equatorial
inner edge.

Two-way structures of this kind may arise astrophysically when a
retrograde inflow first impinges on the outlying parts of a
pre-existing co-rotating accretion disk, and may survive on the order
of an accretion time. (At least in principle, one might conceive of
steady-state configurations in which magnetic coupling of the disk to
the forward-spinning hole supplies the torque needed to drive the
specific angular momentum $\ell$ from negative to positive as the gas
slowly spirals in. By itself, the Rayleigh criterion would forbid
such a transition, since it requires $|\ell|$ to increase inwards for
retrograde motion in the outer rotosphere. However, Archimedean
buoyancy due to the denser inner parts of the disk can easily offset
this. For compressible fluids, H\o iland's criterion is the one that
is relevant (see Blaes 1987).)

The inner parts of disks and tori are constrained by stability and
Bernoulli's law, and must stop well short of the corresponding (co-
or counter-rotating) photon orbits.

In thin (pressureless) disks, all orbits are Keplerian, and the
innermost orbit cannot lie within the last stable Kepler orbit:
$r=r_s$. Stability requires that the Keplerian angular momentum
$|\ell_K|$ should increase outward. From
\begin{equation}
\ell=\rho^2\om=\rho v/(1-v^2)^{\f12}
\end{equation}
one sees that $|\ell_K|$ becomes infinite at both ends of the range
$r_{ph}<r<\infty$, and hence must attain a minimum at a radius
$r_s>r_{ph}$, determined from
\begin{equation}
\pa(\ell_K^2)/\pa r\big|_{r=r_s}=0.
\end{equation}

For the special case of the Kerr geometry,
simple explicit expressions are obtainable for equatorial circular
orbits in both pro- and retrograde cases, and are summarized in
the next section and Appendix~B.

In the case of ``thick disks'' (tori), internally supported by fluid
pressure, material at the inner equatorial edge follows a nearly
Keplerian orbit subject to the condition that its binding energy,
$(1-\vep_K)$, be positive (Kozlowski, Jaroszy\'nski \& Abramowicz
 1978; see also
Appendix~A). Here, the specific energy $\vep$ (energy per unit proper
mass) is given by
\begin{equation}
\vep=-u_\a\xi_{(t)}^\a=\f12\,\pa L/\pa \dot
t=(1-v^2)^{-\f12}(V+\rho\Om_B v).
\end{equation}

For pro- and retrograde Keplerian orbits around non-extremal holes,
$\vep_{+K}$ and $\vep_{-K}$ drop from infinity as one moves outward
from the corresponding photon orbits at $r_{+ph}$ and $r_{-ph}$,
attain minima at radii $r_{+s}$ and $r_{-s}$ respectively and finally tend
to unity from below as $r\to\infty$. It follows that there exist
radii $r_{\epsilon b}$ ($\epsilon=\pm$) such that $r_{\epsilon
ph}<r_{\epsilon b}<r_{\epsilon s}$ and  $\vep_{\epsilon K}(r_{\epsilon
b})=1$. Inner edges of pro- and retrograde thick disks cannot fall
within the corresponding radii of zero binding energy $r_{+b}$ and
$r_{-b}$. This condition is less restrictive than that for thin disks.

\section{Circular orbits in Kerr geometry}

The formulae and results derived in the previous section for orbits
on general stationary and axisymmetric fields take a rather simple
explicit form in the case of the Kerr geometry. The explicit formulae
should be a satisfactory approximation when the mass of the accreting
disk or torus is small. We collect the most interesting results here;
more detail can be found in Appendix~B (also Bardeen 1973,
Lynden-Bell 1978, de~Felice \& Usseglio-Tomasset 1991, de~Felice 1994).

We work in standard (Boyer-Lindquist) co-ordinates for a Kerr
geometry of mass $m$ and angular momentum $ma$. It is convenient to
introduce dimensionless quantities $\xi$, $a_*$ defined by
\begin{equation}
r=m\xi^2,\qquad a=ma_*.
\end{equation}

For equatorial circular orbits, the pro- and retrograde Keplerian
3-velocities $v_+$ and $v_-$ relative to a ZAM observer are given by
$$
v_\epsilon=\ep\left(\f{m}{r^2-2mr+a^2}\right)^{\f12}\,\f{r^2+a^2-2\ep
a(mr)^{\f12}}{r^{\f32}+\ep a m^{\f12}}
$$
with $\ep=\pm1$. In terms of the dimensionless quantities (21),
$$
v_\ep=\ep\,\f{\xi^4+a_*^2-2\ep a_*
\xi}{(\xi^4-2\xi^2+a_*^2)^{\f12}(\xi^3+6a_*)}.
$$

The pro- and retrograde photon orbits $v_\ep=\ep$ have radii $r_{\ep
ph}=m\xi_{\ep ph}^2$, where
$$
\xi_{\ep ph}(3-\xi_{\ep ph}^2)=2\ep a_*.
$$
(Another (prograde) solution is $\xi_{+ph}=0$, corresponding to the
singular equatorial ring $r=0$, $\th=\pi/2$ inside the horizon.
Although of no astrophysical relevance, this has interesting
implications for the source structure of the analytically extended
Kerr geometry (Israel 1977).) For the Schwarzschild case ($a_*=0$),
$\xi_{\ep ph}^2=3$, and we recover the familiar result $r_{ph}=3m$
for both pro- and retrograde photon orbits. For maximally rotating
Kerr ($a_*=1$) we obtain $\xi_{-ph}=2$, $\xi_{+ph}=1$: the retrograde
photon circles out at radius $4m$, the prograde one on the horizon
itself, at radius $m$.

The innermost stable Kepler orbits (pro- and retrograde) are at radii
$r_{\ep s}=m\xi_{\ep s}^2$, where
$$
\xi_{\ep s}\big[4-(3\xi_{\ep s}^2-2)^{\f12}\big]=3\ep a_*.
$$
This gives $\xi_{\pm s}^2=6$, i.e., $r_s=6m$ for Schwarzschild. For
maximal Kerr, $\xi_{-s}=3$, $\xi_{+s}=1$. Counter-rotating thin disks
have inner edges beyond $r_{-s}=9m$, far outside the outer
rotosphere, which terminates at $r_{-ph}=4m$. On the other hand,
co-rotating thin disks can extend within the ergosphere, almost to
the horizon.

Keplerian orbits have energy per unit proper mass
\begin{equation}
\vep_{\ep K}=\f{\xi^3-2\xi+\ep a_*}{\xi^{\f32}(\xi^3-3\xi+2\ep
a_*)^{\f12}}.
\end{equation}
Orbits of zero binding energy, $1-\vep_{K\ep}=0$, which arc believed
to fence off the inner edges of thick disks (see Kozlowski et al.
1978 and Appendix~A) are at radii $r_{\ep b}=m\xi_{\ep b}^2$, where
$$
\xi_{\ep b}(2-\xi_{\ep b})=\ep a_*.
$$
For Schwarzschild, $r_b=4m$. For retrograde orbits in extremal Kerr,
$r_{-b}=(3+2\sqrt 2)m$. Prograde orbits never reach zero binding in
extremal Kerr; evaluation of (22) gives
$$
1-\vep_{+K}=1-3^{-\f12}=0.423\quad{\rm for}\  a_*=1,\ r=m,
$$
a well-known result (Bardeen 1970b), which gives the maximum energy
extractible from accretion onto a Kerr black hole.

At the radius $r_{-ph}$ of the retrograde photon orbit, the specific
energy of the prograde Kepler orbit is
$$
\vep_{+K}(r_{-ph})=\f1{2\sqrt 2}\,\f{3\xi^2-7}{\xi(\xi^2-3)^{\f12}},
\qquad \xi(\xi^2-3)=2a_*.
$$
This gives a positive binding energy for $(a/m)>8\sqrt 2-11=0.3137$.
If the hole is spinning faster than this (very moderate) value, a thick
disk whose outer parts are retrograde can extend into the outer
rotosphere provided its inner edge is co-rotating. That such hybrid
objects can play more than a transitory role in the fuelling of quasar
activity by retrograde accretion is a priori perhaps unlikely, but
really deserves further study.

\section{Concluding remarks}

We hope that the simple treatment we have presented will contribute
to the understanding of the general-relativistic dynamics of the
inner parts of accretion disks near spinning black holes, in
particular, the effects of frame-dragging, which, as we have seen, is
very different for co-rotating and counter-rotating disks. Recent
observational evidence suggests that retrograde accretion may occur
sometimes or even fairly often in galactic nuclei.

We have touched in passing on an issue which has recently had
considerable exposure: Are the paradoxical effects inside circular
photon orbits best ``explained'' as a reversal of centrifugal force,
or is it preferable to suppose that kinetic energy has weight?
Heuristic   explanations are a subjective matter in which each
individual is free to prescribe for himself. Simplicity (a purely
subjective criterion) will be the deciding factor in each case. An
explanation so subtle that it needs explaining loses its raison
d'\^etre. We cede the last word to Richard Feynman:

{\narrower\smallskip
``What do I mean by understanding? Nothing deep or accurate---just to
be able to see some of the qualitative consequences of the equations
without solving them in detail.'' (Feynman 1947)\smallskip}

\section*{Acknowledgements}

One of us (WI) is indebted to Don Page for first
 arousing his interest in this subject
in the summer of 1993. We are grateful to him
for discussions and comments.

This research was supported by the Canadian Institute for Advanced
Research, CNRS and Minist\`ere de la Recherche et Technologie
(France) and NSERC of Canada.

\makeatletter
\renewcommand{\@biblabel}[1]{#1}
\makeatother

\setcounter{equation}{0}
\appendix
\renewcommand{\theequation}{\thesection\arabic{equation}}
\section{Circular orbits in stationary axisymmetric geometry:
covariant approach}

For any steady motion in a circle (not necessarily equatorial)
centred on the symmetry axis, the normalized 4-velocity is
\begin{equation}
u^\a=U^{-1}U^\a,\qquad U^\a=\xi_{(t)}^\a+\Om \xi_{(\vp)}^\a,
\end{equation}
with
\begin{equation}
-U^2=U_\a U^\a=g_{\vp\vp}\Om^2+2\Om g_{\vp t}+g_{tt}.
\end{equation}
Axial symmetry and stationarity imply
\begin{equation}
u^\beta \pa_\beta U=u^\beta \pa_\beta \Om=0.
\end{equation}

The force on a unit mass in this orbit is
$$
F_\a=u_{\a|\beta} u^\beta=U^{-2} U_{\a|\beta}U^\beta.
$$
Substituting from (A1) for $U_{\a|\beta}$ and using (A3) and
Killing's equations $\xi_{\a|\beta}=-\xi_{\beta|\a}$ for both Killing
vectors, we easily reduce this to
$$
F_\a=-U^{-2}(U_{\beta|\a}-\xi_{(\vp)\beta}\pa_\a \Om)U^\beta,
$$
or, by (A2),
\begin{equation}
F_\a=U^{-1}(\pa_\a U+\ell \pa_\a\Om),
\end{equation}
where $\ell=u_\a\xi_{(\vp)}^\a$ is the specific angular momentum.

A ZAM orbit is defined by $\ell=0$. Then $U_{{\rm ZAM}}=V$ and (A4)
reduces to (16) in the text.

The energy per unit rest-mass is
\begin{equation}
\vep=-u^\a
\xi_{(t)\a}=-u^\a(Uu_\a-\Om\xi_{(\vp)\a})=U+\Om\ell,
\end{equation}
where we have used (A1). For a ZAM orbit,
\begin{equation}
\vep_{{\rm ZAM}}=-u_{{\rm ZAM}}^\a \xi_{(t)\a}=U_{{\rm
ZAM}}=V.
\end{equation}

 From (A1) and its analogue for $U_{{\rm ZAM}}^\a$ we find
\begin{equation}
U\ell=U^\a\xi_{(\vp)\a}=(U^\a-U^\a_{{\rm
ZAM}})\xi_{(\vp)\a}=(\Om-\Om_B)g_{\vp\vp}.
\end{equation}
Similarly, subtracting its ZAMO analogue from (A2) and recalling
$\Om_B=-g_{\vp t}/g_{\vp\vp}$,
$$
V^2-U^2=g_{\vp\vp}(\Om-\Om_B)^2.
$$

The relative 3-velocity $v$ of the particle and a ZAM observer in the
same spatial orbit is given by
\begin{equation}
(1-v^2)^{-\f12}=-u_\a u_{{\rm ZAM}}^\a=-U^{-1}\xi_{(t)\a}u^\a_{{\rm
ZAM}}=V/U,
\end{equation}
where we have used (A1) and (A6). From (A8) and (A7),
\begin{equation}
v=V^{-1} (g_{\vp\vp})^{\f12} (\Om-\Om_B).
\end{equation}

(A5) allows us to write (A4) in the alternative form
\begin{equation}
F_\a=U^{-1}(\pa_\a \vep-\Om\pa_\a\ell).
\end{equation}

All these formulae of course extend at once to a continuous medium in
steady rotation, with energy per unit rest-mass
$$
\vep=-u_\a \xi_{(t)}^\a=U+\Om\,\ell=V(1-v^2)^{-\f12}+\Om_B\ell
$$
and angular momentum per unit rest-mass
$$
\ell=u_\a \xi_{(\vp)}^\a=(g_{\vp\vp})^{\f12}\,v/(1-v^2)^{\f12}.
$$
These formulae agree with (18) and (20) in the case of equatorial orbits.

If the medium is a fluid, with local energy density $\mu$ and
pressure $P$, the force is provided by the pressure gradient:
$$
F_\a=\pa_\a P/(\mu+P).
$$
In a ``barytropic'' fluid, $\mu$ depends only on $P$. Then $F_\a$ is
a gradient, and we have the usual stringent consequences exemplified
by von~Zeipel's theorem (Eddington 1926). In particular, the free
surface $(P=0)$ of a barytropic fluid torus is also a surface of
constant specific energy: $\vep=\vep_0$. If the surface is closed,
fluid at the surface must have positive binding energy $(\vep_0<1)$,
since the fluid orbit at the outer equatorial edge is Keplerian or
sub-Keplerian. (On the verge of accretion, the {\it inner\/}
equatorial edge is expected to be cusped and slightly
super-Keplerian, see Kozlowski et al. 1978.)

To conclude, let us briefly link up the present approach with the
force formula (15) for equatorial circular orbits.

Axi-stationarity and (A4) imply generally that $F_\vp=F_t=0$. For an
equatorial orbit, the poloidal component is zero also, and the force
is purely radial. For a specific orbit we can take $\Om$ to be
a given
(position-independent) number. (The apparent dependence on
$\nabla\,\Om$ in (A4) is actually illusory, because it is cancelled
by a part of $\nabla U$.) Then (A4) reduces to the scalar relation
$$
F=\f1{2U^2}\,\f{d}{ds_r}(U^2)=\f1{2V^2(1-v^2)}\,\f{d}{ds_r}(U^2)
$$
where $ds_r$ is the element of proper radial distance, and we have
used (A8).

According to (A2), $U^2$ depends quadratically on $\Om-\Om_B$, which
in turn is simply proportional to $v$ according to (A9); all
coefficients are purely geometrical, i.e., velocity-independent.
Putting all this together, we arrive at
\begin{eqnarray}
F&=-\f1{2V^2(1-v^2)}\,\f{dg_{\vp\vp}}{ds_r}\,\f{V^2}{g_{\vp\vp}}
\,(v-v_+)(v-v_-)\nonumber \\
&=-\f1{2g_{\vp\vp}}\,\f{dg_{\vp\vp}}{ds_r}\,\f{(v-v_+)(v-v_-)}{(1-v^2)},
\end{eqnarray}
in agreement with (15). (A11) itself identifies the quadratic roots
$v_+$, $v_-$ as Keplerian orbital velocities, for which $F=0$.

\section{More on equatorial Kerr orbits}

In Boyer-Lindquist co-ordinates, the Kerr equation $\th=\pi/2$ has a
metric of the form (8), i.e.,
$$
ds^2=\eta^2 dr^2+\rho^2(d\vp-\Om_B dt)^2-V^2\,dt^2,
$$
with
$$
\rho^2=q(r)/r,\qquad \rho^2 V^2=\Delta(r),\qquad \Om_B=2ma/q(r)
$$
and
$$
q(r)=r^3+a^2r+2ma^2,\qquad \Delta(r)=r^2-2mr+a^2.
$$

The quantities defined in (10) take the explicit form
$$
\displaylines{\a=(r^3-ma^2)/qr,\qquad
\beta=ma(3r^2+a^2)/qr\Delta^{\f12}\cr
\gamma=m[(r^2+a^2)^2-4ma^2r]/qr\Delta\cr}
$$

To compute $\delta$, it helps to note the factorization
$$
[(r^2+a^2)^2-4ma^2r](r^3-ma^2)+ma^2(3r^2+a^2)^2=rq^2.
$$
Then it easily follows that
$$
\delta=(m/r\Delta)^{\f12}.
$$

Substituting these expressions into (9) produces the formulae given
in Sec.~4 for the Keplerian velocities $v_{\pm}$. The Keplerian
energies $\vep_{\pm K}$ are then found from (20). It helps to note
the factorization
$$
(1-v_\ep^2)^{\f12}=\f{\rho\xi^{\f32}}{m(\xi^3+\ep
a_*)}\left(\f{\xi^3-3\xi+2\ep a_*}{\xi^4-2\xi^2+a_*^2}\right)^{\f12}
$$
in the notation of Sec.~4.

The Keplerian angular momenta, prograde and retrograde, are
$$
\ell_\ep=\rho v_{\ep}(1-v_\ep^2)^{-\f12}=m\ep\,\f{\xi^4+a_*^2-2\ep
a_*\xi}{\xi^{\f32}(\xi^3-3\xi+2\ep a_*)^{\f12}}.
$$
The condition for marginal Keplerian stability, $\pa(\ell^2)/\pa
r=0$, factorizes as
$$
(\xi^3+\ep a_*)(\xi^4-6\xi^2+8\ep a_*\xi-3a_*^2)=0.
$$
The vanishing of the second factor is most easily treated as a
quadratic equation for $a_*$ in terms of $\xi$, and leads to the
results stated in Sec.~4.
\end{document}